\documentstyle[12pt]{article}

\newcommand{\be}{\begin{equation}}
\newcommand{\ee}{\end{equation}}
\newcommand{\bea}{\begin{eqnarray*}}
\newcommand{\eea}{\end{eqnarray*}}
\newcommand{\ba}{\begin{eqnarray}}
\newcommand{\ea}{\end{eqnarray}}

\begin{document}
\begin{titlepage}
\begin{flushright}
{\large \bf UCL-IPT-98-01}
\end{flushright}
\vskip 2cm

\begin{center}
{\Large {\bf Final State Interaction Phases in $(B \to K \pi)$ Decay
Amplitudes} }

\vskip .4in

{\large {\it D. Del\'epine, J.-M. G\'erard, J. Pestieau, J. Weyers } }\\[.15in]
\vskip 2cm
{\em Institut de Physique Th\'eorique\\
 Universit\'e catholique de Louvain\\
B-1348 Louvain-la-Neuve, Belgium}

\end{center}

\vskip 2in

\begin{abstract}

A simple Regge pole model for  $K\pi$ scattering explains
the large  phase  $e^{i \delta}$ between isospin
amplitudes which is observed at the $D$ meson mass $(\delta \approx \frac{\pi%
}{2})$. It  predicts $\delta \approx 14^\circ - 20^\circ$ at the $B$ mass.
Implications for $(B \to K \pi)$ decays and extensions of
the model to other two-body decay channels are briefly discussed.
\end{abstract}
\end{titlepage}

\newpage

\renewcommand{\thepage}{\arabic{page}}
\setcounter{page}{1}
\setcounter{footnote}{1}

\section{Introduction}

\par \ \ \ \ With $B$-factories forthcoming, detailed checks of the precise CP-violation
pattern predicted by the standard model will become possible. However it is
by no means trivial to extract reliable information on CP-violation
parameters from various $B$-decay modes. One of the problems is of course
how to estimate ``hadronic effects" such as final state interaction (FSI)
phases. Although these phases are of no particular interest by themselves,
they do play an important role for many potential signals of CP violation in
hadronic $B$-decays.

The relevant question concerning these FSI phases is whether they are
significantly different from 1 or not. Clearly the answer to this question
depends on the hadronic channels considered. Here we will focus our
attention on $(K \pi)$ channels where experimental data also exist for $B$ decays\cite{1}. There are two isospin invariant scattering
amplitudes $(I = 1/2$ and $I = 3/2$) and the quantity one wants to estimate, 
as a function of energy, is $\delta (s) = \delta_3 (s) - \delta_1 (s)$
namely the difference between the $S$-wave phase shifts in the $I = 3/2 $ \ $
(\delta_3)$ and $I = 1/2 $ \  $(\delta_1)$ amplitudes. As a matter of fact $\delta
(s)$ has been measured at the $D$ mass $(s = m^2_D)$ where it is
found\cite{2} to be around $\pi/2$. Naively one does not expect such a
huge FSI angle at $s = m^2_D$ to become negligible at $s = m^2_B$ but,
obviously, a more quantitative argument is called for.

The main purpose of this letter is to suggest a Regge model as  a
general strategy for determining FSI angles\cite{3}.
Past experience with $\pi N$ and $\bar K N$ scattering amplitudes strongly
suggests that such a model should work quite well for $K\pi$ scattering
over an energy range which includes the $D$ and $B$ meson masses.

The dominant Regge exchanges to consider in $K \pi \to K \pi$ scattering
are, respectively, the Pomeron $(P)$ and the exchange degenerate $\rho-f_2$
trajectories in the $t$-channel and in the $u$-channel the exchange degenerate $K^\ast -
K^{\ast\ast}$ trajectories. In the next section we
briefly recall a few properties of these trajectories and then proceed to
show in section 3 that with all parameters fixed phenomenologically our
model automatically accounts for the observed $\delta (m^2_D) \simeq \frac{%
\pi}{2}$. From the known energy dependence of Regge trajectories one then
readily predicts $\delta (m^2_B) $ close to  $20$ degrees, namely  quite
 a sizeable FSI angle at the $B$ mass, as na\"\i vely expected. These are our main
results.

To conclude this note we first comment on obvious implications of our
results for  $(B \to K \pi)$ decays and then end with
several general remarks on the parametrization of any (quasi) two-body decay
amplitude of the $B$ mesons.

\section{A Regge model for $(K\pi \to K \pi)$ scattering amplitudes}

\par \ \ \ \  We take $s, t, u$ to be the usual Mandelstam variables. In the $s$-channel, $%
(K \pi \to K \pi)$ scattering amplitudes are linear combinations of the
isospin invariant amplitudes $A^s_{1/2}$ and $A^s_{3/2}$. In the $t$-channel 
$(K \overline K \to \pi \pi)$, we have isospin invariant amplitudes $A^t_0$
(isospin 0) and $A^t_1$ (isospin 1) and, similarly, in the $u$-channel $(K
\pi \to \pi K)$, we define $A^u_{1/2}$ and $A^u_{3/2}$. The relations between
these amplitudes are given by the crossing matrices 
\begin{equation}
\left( 
\begin{array}{c}
A^s_{1/2} \\ 
A^s_{3/2}
\end{array}
\right) = \left( 
\begin{array}{cc}
\frac{1}{\sqrt{6}} & 1   \\ 
\frac{1}{\sqrt{6}} & - \frac{1}{2}  
\end{array}
\right) \ \ \left( 
\begin{array}{c}
A^t_0 \\ 
A^t_1
\end{array}
\right) = \left( 
\begin{array}{cc}
1/3 & 4/3 \\ 
-2/3 & 1/3  
\end{array}
\right) \left( 
\begin{array}{c}
A^u_{1/2}   \\ 
A^u_{3/2}  
\end{array}
\right).
\end{equation}

In a Regge model, $s$-channel amplitudes at high energy (large s) are
parametrized as sums over (Regge pole) exchanges in the crossed channels:
near the forward direction ($t$ small), $t$-channel exchanges dominate while
near the backward direction ($t$ close to $-s$ or $u$ small), it is the $u$%
-channel exchanges which are relevant.

The generic form, at large $s$, of a Regge pole exchanged in the $t$-channel
is given by 
\begin{equation}
-\beta (t) \frac{\tau + e^{-i \pi \alpha (t)}}{\sin \pi \alpha (t)} \left( 
\frac{s}{s_o} \right)^{\alpha (t)}.
\end{equation}
In Eq.(2), $\beta (t)$ is the residue function, $\tau$ the signature $(\tau = \pm 1)$
and 
\begin{equation}
\alpha (t) = \alpha_o + \alpha^{\prime}t
\end{equation}
is the (linear) Regge trajectory with intercept $\alpha_o$ and slope $%
\alpha^{\prime}$; finally $s_o$ is a scale factor usually taken as 1 Gev$^2$%
. For a Regge pole exchanged in the $u$-channel, the generic form is similar
to Eq.(2) but with the variable $t$ replaced by $u$.

The leading trajectory (highest intercept) is the so-called Pomeron $(P)$.
It has the quantum numbers of the vacuum $(I=0,\tau=+1$) and its exchange
describes ``diffractive scattering". The Pomeron always contributes to
elastic scattering and describes quite well the bulk of hadronic
differential cross-sections over a wide energy range.

In the energy interval which is of interest to us here, namely 
\begin{equation}
3 \mbox{ Gev}^2 {<\hspace{-3.5mm}\raisebox{-1mm}{$\scriptstyle \sim$ }} s {\hspace{3mm}}  {<%
\hspace{-3.5mm}\raisebox{-1mm}{$\scriptstyle \sim $ }} 35 \mbox{ Gev}^2
\end{equation}
a very simple but excellent phenomenological parametrization of the Pomeron
trajectory and residue function is given by 
\begin{equation}
\alpha_P (t) = 1 \ \ \ \ \ \ \ \ \ \ \ 
\end{equation}
and
\begin{equation}
\beta_P (t) = \beta_P (0) e^{b_P t}
\end{equation}
with 
\begin{equation}
2.5 \mbox{ Gev}^{-2} {<\hspace{-3.5mm}\raisebox{-1mm}{$\scriptstyle \sim$ }}
b_P {\hspace{3mm}}  {<\hspace{-3.5mm}\raisebox{-1mm}{$\scriptstyle \sim$ }} 3 \mbox{ Gev}^{-2}
\end{equation}
obtained from fits\cite{4} to elastic $\pi p$, $pp$ and $Kp$ differential cross-sections 
(using factorization).
As a result the Pomeron contribution to $A^t_0$ now reads $(s_o =$ 1 Gev$^2$%
) 
\begin{equation}
A_P = i \beta_P (0) e^{b_P t} s.
\end{equation}

The next trajectories to consider are the $\rho-f_2$ trajectories in the $t$%
-channel and the $K^\ast-K^{\ast\ast}$ trajectories in the $u$-channel. The $%
\rho$ trajectory has $T=1, \tau=-1$ while the $f_2$ trajectory has $%
T=0,\tau=+1$; similarly the $K^\ast$ trajectory has $T=1/2,\tau=-1$ while
the $K^{\ast\ast}$ trajectory has the opposite signature. Because of the
absence of exotic resonances (no $K\pi$ resonances with $I=3/2$), the $\rho$
and $f_2$ trajectories as well as the $K^\ast-K^{\ast\ast}$ ones must be
exchange degenerate. Specifically this means that the $\rho$ and $f_2$
trajectories co\"\i ncide  
\begin{equation}
\alpha_\rho (t) = \alpha_{f_2} (t) \cong \frac{1}{2} + t
\end{equation}
and that their residues are related i.e.
\begin{equation}
\frac{\beta_{f_2} (t)}{\sqrt{6}} = \frac{\beta_\rho (t)}{2}. \ \ \ \ \ \ \ \
\ \ \ \ 
\end{equation}

Similarly, for the $K^\ast-K^{\ast \ast}$ trajectories (in the $SU(3)$%
-limit) 
\begin{equation}
\alpha_{K^\ast} (u) = \alpha_{K^{\ast\ast}} (u) \cong \frac{1}{2} + u
\end{equation}
and
\begin{equation}
-\beta_{K^\ast} (u) = \beta_{K^{\ast\ast}} (u).
\end{equation}
Eqs(9-10) and Eqs(11-12) guarantee that the non diffractive imaginary part
of $A^s_{3/2}$ vanishes. They used to be called ``duality constraints"\cite{weyers}.

We neglect lower lying trajectories such as the $\rho^{ \prime}(I=1,\tau=-1)$
and the $f_0 (I=0,\tau=+1)$ in the $t$-channel as well as their $SU(3)$
partners in the $u$-channel. Were we to include them they should also be
taken as exchange degenerate.

It is customary to write the residue function of the $\rho$ trajectory as 
\begin{equation}
\beta_\rho (t) = \frac{\overline{\beta}_\rho (t)}{\Gamma (\alpha(t))} .
\end{equation}
 Since $\Gamma (\alpha (t)) \sin \pi \alpha (t)$ is a very smooth function
of $t$, no harm is done in using at small $t$ the approximations  
\begin{eqnarray}
\Gamma (\alpha (t)) \sin \pi \alpha (t) \approx \Gamma (\alpha (0)) \sin \pi
\alpha (0) = \sqrt{\pi} \\
\overline{\beta}_\rho (t) \approx \overline{\beta}_\rho (0) \hspace*{30mm}
\end{eqnarray}
and in writing the $\rho$ trajectory contribution to $A^t_1$ as 
\begin{equation}
A_\rho (s, \ \mbox{small} \ t) = \frac{\overline{\beta}_\rho (0)}{\sqrt{\pi}}
\Bigl(1+i \exp (-i\pi t)\Bigr) s^{0.5+t}.
\end{equation}
An exactly similar reasoning gives for the $f_2$ trajectory contribution to $%
A^t_0$ 
\begin{equation}
A_{f_2} (s, \ \mbox{small} \ t) = \sqrt{\frac{3}{2}} \frac{\overline{\beta}%
_\rho (0)}{\sqrt{\pi}} \Bigl(-1 + i \exp (-i \pi t) \Bigr) s^{0.5 + t}
\end{equation}
while for the $K^\ast$ and $K^{\ast\ast}$ trajectories contributions to $%
A^u_{1/2}$ one writes 
\begin{equation}
A_{K^\ast} (s, \ \mbox{small} \ u) =  \frac{\overline{\beta}_{K^\ast} (0)}{%
\sqrt{\pi}} \Bigl(1+i \exp (-i\pi u)\Bigr) s^{0.5 + u}
\end{equation}
\begin{equation}
A_{K^{\ast\ast}} (s, \ \mbox{small} \ u) = - \frac{\overline{\beta}_{K^\ast}
(0)}{\sqrt{\pi}} \Bigl(-1+i \exp (-i\pi u)\Bigr) s^{0.5 + u}
\end{equation}
with 
\begin{equation}
\overline{\beta}_{K^\ast} (0) = \frac{3}{4} { } \overline{\beta}_\rho (0)
\end{equation}
in the $SU(3)$ limit.

Putting everything together and using the crossing matrices given in Eq.(1), our
Regge model for $K\pi$ scattering is now completely defined by the amplitudes 
$$
A^s_{1/2} (s, \ \mbox{small} \ t) = \frac{i}{\sqrt{6}} \beta_P (0) e^{b_P t}
s + \frac{\overline{\beta}_\rho (0)}{2\sqrt{\pi}} s^{0.5 + t} + \frac{3 i 
\overline{\beta}_\rho (0)}{2\sqrt{\pi}} e^{-i \pi t} s^{0.5 + t} \eqno{(21a)}
$$
$$
A^s_{1/2} (s, \ \mbox{small} \ u) = \frac{\overline{\beta}_\rho (0)}{2 \sqrt{%
\pi}} s^{0.5 + u} \hspace*{64mm} \eqno{(21b)} 
$$
and 
$$
A^s_{3/2} (s, \ \mbox{small} \ t) = \frac{i}{\sqrt{6}} \beta_P (0) e^{b_P t}
s - \frac{\overline{\beta}_\rho (0)}{\sqrt{\pi}} s^{0.5 + t} \eqno{(22a)} 
$$
$$
A^s_{3/2} (s, \ \mbox{small} \ u) = - \frac{\overline{\beta}_\rho (0)}{ 
\sqrt{\pi}} s^{0.5 + u} \hspace*{27mm} \eqno{(22b)} 
$$

\section{S-wave Rescattering Phases}

\setcounter{equation}{22}

\par \ \ \ \  The remaining task is now to extract from Eqs(21-22) the $\ell = 0$ partial
wave amplitudes $a_{1/2} (s)$ and $a_{3/2} (s)$. Neglecting $\pi$ and $K$ masses, 
we have, up to irrelevant real factors
\begin{equation}
a_I (s) \propto \int^0_{-s} dt A^s_I (s,t).
\end{equation}

\par From the physical ideas underlying Eqs(21-22) it is clear that outside the
forward and backward regions, the integral in Eq.(23) gives a negligibly small
contribution to $a_I (s)$. We thus write 
\begin{equation}
a_I (s) \propto \int^0_{t_o} dt A^s_I (s, \ \mbox{small} \ t) + \int^0_{u_o}
du A^s_I (s, \ \mbox{small} \ u).
\end{equation}

With the explicit expressions given in Eqs(21-22), the integrals in Eq.(24)
are trivial to perform. Furthermore, the integrated contributions at the $t_o$
and $u_o$ boundaries (around $1$ GeV$^2$) are considerably smaller than at the boundary 
$0$ of both integrals in Eq.(24).
Neglecting these contributions, one thus obtains 
\begin{equation}
a_{1/2} (s) = \frac{i}{\sqrt{6}} \frac{\beta_P (0)}{b_P} s + \frac{\overline{%
\beta}_\rho (0)}{\sqrt{\pi}} \frac{1}{\ln s} s^{1/2} + \frac{3i}{2\sqrt{\pi}}
\overline{\beta}_\rho (0) \frac{\ln s + i \pi}{(\ln s)^2 + \pi^2} s^{1/2}
\end{equation}
and 
\begin{equation}
a_{3/2} (s) = \frac{i}{\sqrt{6}} \frac{ \beta_P (0)}{b_P} s - 2 \frac{%
\overline{\beta}_\rho (0)}{\sqrt{\pi}} \frac{1}{\ln s} s^{1/2}
\end{equation}
from which the tan$(\delta_I) = \frac{Im \mbox{ } a_{I} (s)}{Re \mbox{ } a_{I}
(s)}$  are straightforward to compute. Note that both tan$
(\delta_1)$ and tan$(\delta_3)$ depend on one single phenomenologically
determined parameter namely 
\begin{equation}
x = \frac{\sqrt{\pi} \beta_P (0)}{b_P \overline{\beta}_\rho (0)}.
\end{equation}

 \par From fits\cite{5} to $\pi p$, $p p$ and $K p$  total cross sections in the energy 
 range given in Eq.(4) (again using factorization), we find 
 \begin{equation}
 \frac{%
\sqrt{\pi} \beta_P (0)}{\overline{\beta}_\rho (0)} = 2.9 \pm 0.2 .
\end{equation}
From Eq.(7), we thus conclude that $x$ is close to one
\begin{equation}
x=1.07 \pm 0.17 .
\end{equation}
Similar results are obtained using the fits given in Ref.\cite{6} for a larger energy range. \\
 \par With these  values for $x$, the range for the FSI angle
 at the $D$ mass is  \underline{calculated} to be
 \begin{equation}
 \delta(m_D^2) \equiv \delta_3(m_D^2)-\delta_1(m_D^2)=(85 \pm 6)^\circ
 \end{equation}
in spectacular agreement with the recent analysis of CLEO data\cite{2}
 \begin{equation}
 \delta(m_D^2)=(96\pm 13)^\circ .
 \end{equation}
We stress that both the analysis of CLEO data and our calculation are based 
on the quasi-elastic approximation.

\par At the $B$ mass, we predict a sizeable angle close to $20$ degrees, namely
\begin{equation}
\delta(m_B^2)=(17\pm 3)^\circ .
\end{equation}

Before commenting on our prediction for $\delta (m^2_B)$, it may be
worthwhile to point out a few facts about our calculation of $\delta (s)$

\begin{enumerate}
\item[-]  it is a no-parameter calculation: $x$ is determined from the data
on total cross-sections\cite{5} and $\frac{d\sigma}{dt}$'s \cite{4};

\item[-]  in performing our calculation of $\delta(s)$, we have made several
approximations a.o. we neglected  lower trajectories as well as the 
intermediate region in the S-wave projection integral. These approximations are certainly sound from
a phenomenological point of view and they become better and better as $s$
increases.  At the $D$ mass we do not believe that our end result should
be trusted to better than 10-20$\%$ but in any case,  agreement with the
data remains excellent;

\item[-]  the calculations presented here for $K\pi$ scattering can of
course be repeated for $\pi\pi$ or $K \bar K$ scattering. A detailed account
and discussion of these calculations will be presented elsewhere\cite{7}. Here we
simply point out that the results of both  calculations are once again
in excellent agreement with the data available\cite{2} at the $D$ mass : $\delta^{\pi\pi}$ is found to be
around $\pi /3$  and  $\delta^{K\bar K}$ around $\pi /6$. These results considerably strenghten our confidence in a simple Regge parametrization of hadronic scattering amplitudes.
\end{enumerate}

\section{Conclusions}

\par \ \ \ \  The main results of this letter are given by Eqs(30-32) and can be
summarized as follows: a Regge model for $K\pi$ scattering explains the
large $S$-wave rescattering phase difference $\delta$ observed at the $D$
meson mass namely $\delta (m^2_D) \approx \frac{\pi}{2}$,  and predicts $%
\delta (m^2_B) \approx 20^\circ$.
\par  Such a sizeable FSI angle at the $B$ 
meson mass  leads to important implications for
 $B \to K \pi$ decays\cite{8}: it invalidates the Fleischer-Mannel bound\cite{9} on the 
 Cabibbo-Kobayashi-Maskawa angle $\gamma$ and implies a potentially large $CP$ 
 asymmetry, $ a$, in ($ B^{\pm} \to K\pi^{\pm}$) decays:
 \begin{equation}
 a \approx 4 (\mbox{sin} \gamma) \% .
 \end{equation}
\par Strong interaction hadronic phases can be parametrized a la
Regge for \underline{any} (quasi) two body decay mode of the $B$ meson ($ \pi\pi,K\bar K $ as already mentioned, but also $\pi \rho,K^\ast \pi,K^\ast
\rho$ etc...).
\par The fact that our quasi-elastic treatment of the scattering amplitudes for $K \pi$,$\pi \pi$ and $K \bar{K}$ agrees so well with the data at the $D$ meson mass is a strong argument for neglecting inelastic effects on hadronic phases.  

In view of the previous comments, a general parametrization for all two-body
decay modes of the $B$ mesons naturally suggests itself. The decay amplitude
can be written as a sum of reduced matrix elements  $\ll B \mid H_W \mid (M_1 M_2),I \gg $ of the effective weak
hamiltonian,  multiplied by the appropriate hadronic  FSI phases $e^{\delta_I}$. 
These reduced  matrix elements  are in
general complex numbers which can be systematically calculated in terms of
tree-level, colour suppressed, penguin, exchange or annihilation quark diagrams.
Of course, no isospin violating "scattering phases" are allowed between these diagrams and furthermore, as already shown elsewhere\cite{8}, classes of diagrams which would na\"\i vely be excluded can reappear due to factors of the type $(1-e^{i \delta_I})$. On the other hand, penguin diagrams can provide an absorptive (i.e. imaginary) component to the reduced matrix elements\cite{10}. But these imaginary parts are very model-dependent and probably quite small. Therefore we suggest\cite{12}, as  a first approximation to simply ignore these "quark phases" whenever the hadronic phases are sizeable. This was assumed in Ref.\cite{8}. This happens to be the case for $B \to K \pi$ decays. 
 
\vspace*{15mm}
\noindent
{\Large\bf Acknowledgements}
\par We are grateful to Christopher Smith and Frank W\"urthwein for useful discussions and comments.

\end{document}